\documentclass[aps,prl,twocolumn,showpacs,groupedaddress]{revtex4}
\usepackage{mathrsfs}
\usepackage{amsmath}
\usepackage{amsfonts}
\usepackage{amssymb}
\usepackage{amsthm}
\usepackage{graphicx}
\usepackage{color}
\usepackage{graphics}
\usepackage{epsfig}
\usepackage{ulem}
\usepackage{subfigure}%
\providecommand{\U}[1]{\protect\rule{.1in}{.1in}}
\begin{document}
\title{A thermodynamic theory for thermal-gradient-driven domain
wall motion}
\author{X.S. Wang}
\author{X.R. Wang}
\email[Corresponding author:]{phxwan@ust.hk}
\affiliation{Physics Department, The Hong Kong University of
Science and Technology, Clear Water Bay, Kowloon, Hong Kong}
\affiliation{HKUST Shenzhen Research Institute, Shenzhen 518057, China}

\begin{abstract}
Spin waves (or magnons) interact with magnetic domain walls (DWs) in a
complicated way that a DW can propagate either along or against magnon flow.
However, thermally activated magnons always drive a DW to the hotter
region of a nanowire of magnetic insulators under a temperature gradient.
We theoretically illustrate why it is surely so by showing that DW entropy
is always larger than that of a domain as long as material parameters do
not depend on spin textures. Equivalently, the total free energy
of the wire can be lowered when the DW moves to the hotter region.
The larger DW entropy is related to the increase of magnon density
of states at low energy originated from the gapless magnon bound states.
\end{abstract}

\maketitle
Manipulation of a magnetic domain wall (DW) in a nanostructure
has attracted much attention due to its application prospects in
logical operations \cite{Cowburn} and data storage \cite{Parkin}.
Moving DWs in a controlled manner is an important issue in those
applications. Magnetic fields via energy dissipation \cite
{Walker,energy,WXS} and electric current via angular momentum transfer
\cite{STT,SZhang,ono} are well-known control parameters for DW motion.
To overcome the Joule heating \cite{ono} in current
driven magnetization reversal, heat itself has recently been proposed
\cite{Bauer} as an efficient control parameter for spin manipulation.
A temperature gradient can generate spin current \cite{sse1,sse2,sse3}
due to electron and/or magnon flow. This thermoelectric phenomenon
of spin current generation is called spin Seebeck effect that has been
experimentally observed through the inverse spin Hall effect \cite{sse1}.
The spin Seebeck effect has also been suggested \cite{Nowak,YT} as
a control parameter for DW manipulation. As spin 1 carriers, magnons
can mediate a spin transfer torque (STT) \cite{magnon} on a magnetic
texture like a DW in a similar way as the electrons do.
It was predicted \cite{Nowak,YT} that a thermal-magnon-driven DW can
propagate along a wire at a high speed, and this prediction was
confirmed in a recent experiment \cite{Wangkl}.

There is little doubt that magnonic STT can drive a DW to move.
In terms of DW propagation direction, the pure magnonic STT predicts
\cite{magnon} a DW moving against magnon propagation direction.
However, a DW may also propagate along magnon flow direction \cite{Klaui,Guo,Kim}.
This is very similar to
electric-current-driven DW motion: A DW propagates along or against
electron flow direction, depending on detailed
spin-orbit interactions and DW types \cite{Miron,Beach,Parkin2}.
It is not clear whether magnon-driven ``wrong" DW propagation
direction shares a similar physics origin as its electron counterpart.
In principle, angular momentum does not dictate DW motion since
its governing dynamics, Landau-Lifshitz-Gilbert (LLG) equation,
does not conserve the total angular momentum when the spin-lattice
and spin-orbital interactions are involved. Nevertheless, all
studies \cite{Nowak,YT,Wangkl} showed that a DW propagates
to the hotter part of a wire under a temperature gradient.
Although this result is consistent with the
STT prediction, magnonic STT cannot be the sole physics behind.
It is thus interesting to ask whether there is a general
thermodynamic principle for thermal-gradient-driven DW motion.
Previous theories \cite{Nowak,YT,Hinzke} are based on magnon
kinetics, multiscale micromagnetic framework as well as spin
model simulations. In this paper, the underneath thermodynamic
principle of thermal-gradient-driven DW motion is revealed.
Due to the magnon bound states, the magnon density of states at
low energy in a DW is larger than that in a domain, resulting
in a larger DW entropy at any temperature. Thus, a DW must
propagate to the hotter part of a wire under a thermal gradient
in order to lower the wire free energy by taking the advantage
of the larger DW entropy. Furthermore, our results explain also
decrease of domain size by heating \cite{floating,domainsize,DWF}.
\begin{figure}
\begin{center}
\includegraphics[width=8.5cm]{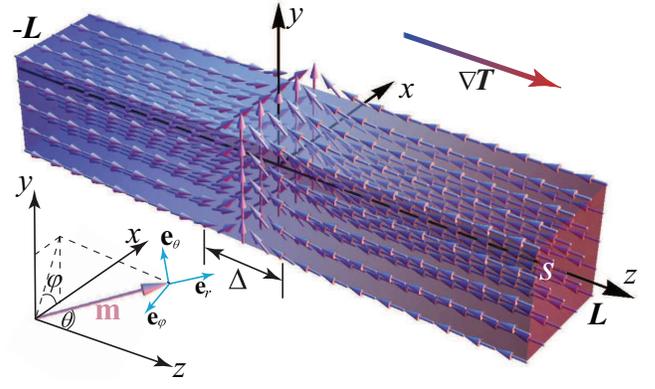}
\end{center}
\caption{(color online) Schematic diagram of a nanowire with a
head-to-head DW at its center ($z=0$) under a temperature gradient
$\nabla{T}$. Pink arrows illustrate magnetization $\mathbf{m}$
and $\mathbf{e}_{r}$, $\mathbf{e}_\theta$, $\mathbf{e}_\phi$
are unit vectors of a spherical coordinates defined in terms of
the $z$-axis and $\mathbf{m}$. The length of the wire is $2L$
and the cross section area is $s$. $\Delta$ is the DW width.
Blue (red) color indicates the colder (hotter) region.}
\label{fig1}
\end{figure}

Magnetic domains and magnetic domain walls are stable and metastable
states of a magnet. At thermal equilibrium, spin textures of both
domains and DWs fluctuate around their energy minimum configurations
at a finite temperature $T$, creating magnons and microscopic states
which contribute to the system entropy and free energy.
To calculate the entropy and free energy of a domain and a DW in a
magnetic nanowire, we consider a head-to-head DW in a bi-axial wire
of length $2L$ along the $z$ direction, as shown in Fig. 1.
The wire is made of a simple cubic crystal of lattice constant $a$.
The static DW structure is assumed to be $z$ dependent only,
independent on $x$ and $y$. The DW is placed at the wire center
and the temperature $T$ is far below the Curie temperature $T_c$.
In the continuous limit of $a\rightarrow 0$, the magnetization is
then governed by the dimensionless LLG equation \cite{LLG},
\begin{equation}
\frac{\partial{\mathbf{m}}}{\partial{t}}=-\mathbf{m}
\times\mathbf{h}_{\text{eff}}+\alpha\mathbf{m}
\times\frac{\partial\mathbf{m}}{\partial{t}},\label{LLG}%
\end{equation}
where $\mathbf{m}$ is the unit direction of magnetization with a
saturation value $M_s$. $t$ is in units of $(\gamma{M_s})^{-1}$
where $\gamma$ is the gyromagnetic ratio. $\alpha$ is the dimensionless
Gilbert damping constant which is negligibly small ($10^{-4}$) for
a magnetic insulator like YIG \cite{YIG}. $\mathbf{h}_\text{eff}=
A\nabla^2\mathbf{m}+K_zm_z\hat{\mathbf{z}}-K_xm_x\hat{\mathbf{x}}$
is the effective field in units of $M_s$ where $A$, $K_z$, and
$K_x$ are exchange constant, the anisotropy constants of the
easy- and hard-axis along the $z$ and $x$ directions, respectively.
The energy density is in units of $\mu_0M_s^2$ so that $K'$s are
dimensionless and $A$ has the dimension of length square \cite{YIG}.
All parameters are assumed to be independent on spin texture and $T$.
The spin wave equation is obtained by linearizing Eq. \eqref{LLG}
for the small fluctuation of the magnetization around either
a domain ($\theta=0$ or $\pi$) or a static DW \cite{Walker} of
$\theta=2\arctan {e^{z/\Delta}}$ and $\phi=\pi/2$, where $\theta$
and $\phi$ are polar and azimuthal angles of
$\mathbf{m}$, and $\Delta=\sqrt{\frac{A}{K_z}}$ is the DW width.
Let $\mathbf{m}\equiv\mathbf{e}_r+\left[m_\theta(x,y,z)\mathbf{e}_\theta
+m_\phi(x,y,z)\mathbf{e}_\phi\right]e^{-i\omega{t}}$, where $\mathbf{e}_r
=\mathbf{m}$ and $\mathbf{e}_\theta$, $\mathbf{e}_\phi$ are defined in
Fig. 1. $\omega$ is spin wave frequency that related to the magnon
energy as ${\varepsilon}=\hbar\omega$. Following Ref. \cite{magnon},
$\psi=m_\theta+icm_\phi$ with $c=\frac{K_x\sqrt{K_x^2+ 4\omega^2}}
{2\omega}$ satisfies the following Schrodinger equation for
spin waves around the Walker DW,
\begin{equation}
q^2\psi(x,y,z)=(-\Delta^2\nabla^2-2\mathrm{sech}
^{2}\frac{z}{\Delta})\psi(x,y,z),
\label{Schrodinger}
\end{equation}%
with $q^2=-1+(\omega c)/K_z$
and $q$ is the norm of wave vector $\mathbf{q}$. The equation has
propagating solutions \cite{magnon,spinwave}:
\begin{equation}
\psi_1=\frac{A_1}{\sqrt{1+(\Delta{q_z})^2}}(-i\Delta q_z+\tanh\frac{z}
{\Delta})e^{i\mathbf{q\cdot x}},
\label{continuous}
\end{equation}
with dispersion relation
\begin{equation}
\omega_1(\mathbf{q})=\sqrt{(Aq^2+K_z)(Aq^2+K_z+K_x)}
\label{dispersion}
\end{equation}
where $A_1$ is the spin wave amplitude of wave vector $\mathbf{q}$,
not confused with the exchange constant. Spectrum \eqref{dispersion}
is gapped with a gap of $\varepsilon_0=\hbar\sqrt{K_z(K_z+K_x)}$.
Eq. \eqref{Schrodinger} has also bound (in $z$ direction) states,
\begin{equation}
\psi_2=\frac{A_2}{\sqrt{2\Delta}}\mathrm{sech}\frac{z}{\Delta}e^{i(q_xx+q_yy)},
\label{bound}
\end{equation}
with amplitude $A_2$ and a gapless spectrum of $\omega_2(q_x,q_y)=
\sqrt{A(q_x^2+q_y^2)[A(q_x^2+q_y^2)+K_x]}$ \cite{spinwave}.

The wave components in transverse directions give a trivial
factor of $s/(2\pi)^2$ to the magnon density of states (DOS)
in $q_x$ and $q_y$, where $s$ is the cross section area.
To find allowed $q_z$ for $\psi_1$, we use the anti-periodic
boundary condition $\psi_1(x,y,-L)=-\psi_1(x,y,L)$
that gives $e^{i(q_zL-\eta)}=-e^{i(-q_zL+\eta)}$
with $\eta=\arctan(\Delta{q_z})$ for $L\gg\Delta$. Thus $q_z'$s
are $q_zL-\eta=-q_zL+(2n+1)\pi+\eta$ with $n=0,\pm 1,\ldots $.
Therefore, the propagating spin waves contribute to 1D magnon DOS
in $q_z$ by
\begin{equation}
\rho_1(q_z)=\frac{dn}{dq_z}=
\frac{L}{\pi}-\frac{\Delta}{\pi[1+(\Delta q_z)^2]}.
\label{dosqz}
\end{equation}
This differs from the domain magnon DOS, $\rho_\mathrm{D}(q_z)=L/\pi$.
The integral of the second term in the right hand side of
Eq. \eqref{dosqz} over $q_z\in[-\frac{\pi}{a},\frac{\pi}{a}]$ is 1
when $\Delta/a\rightarrow\infty$, consistent with the continuous
limit for Eqs. \eqref{LLG}--\eqref{dosqz}. The disappearance of one
propagating spin wave is compensated by one localized spin wave
(bound magnons) of Eq. \eqref{bound} for given $q_x$ and $q_y$.
As required, the total number of spin waves (magnon modes)
inside the Brillouin zone does not change. The upper bound of energy
${\varepsilon}$ of propagating magnon is ${\varepsilon}_c=\hbar\omega_1$
at $q^2=\frac{3\pi^2}{a^2}$. After knowing the distribution of states in
$\mathbf{q}$ space and the dispersion relation, the DOS
$\rho_1({\varepsilon})$ in ${\varepsilon}$ due to propagating spin
waves in a DW can be calculated in a straight forward way,
\begin{equation}
\rho_1({\varepsilon})=\int_{\mathrm{B.Z.}}\frac{s}{(2\pi)^2}\rho_1(q_z)
\delta(\varepsilon-\hbar\omega_1)dq_xdq_ydq_z,
\end{equation}
where B.Z. stands for the Brillouin zone of
$q_x,q_y,q_z\in[-\frac{\pi}{a},\frac{\pi}{a}]$.
The bound states $\psi_2$ also contribute to the density of states by
\begin{equation}
\rho_2(\varepsilon)=\iint\frac{s}{(2\pi)^2}
\delta(\varepsilon-\hbar\omega_2)dq_xdq_y
\label{bdos}
\end{equation}
for $q_x,q_y\in[-\frac{\pi}{a},\frac{\pi}{a}]$.
The total magnon DOS in a DW $\rho_\mathrm{DW}$ is the sum of
$\rho_1(\varepsilon)$ and $\rho_2(\varepsilon)$.

\begin{figure}
\begin{center}
\includegraphics[width=8.5cm]{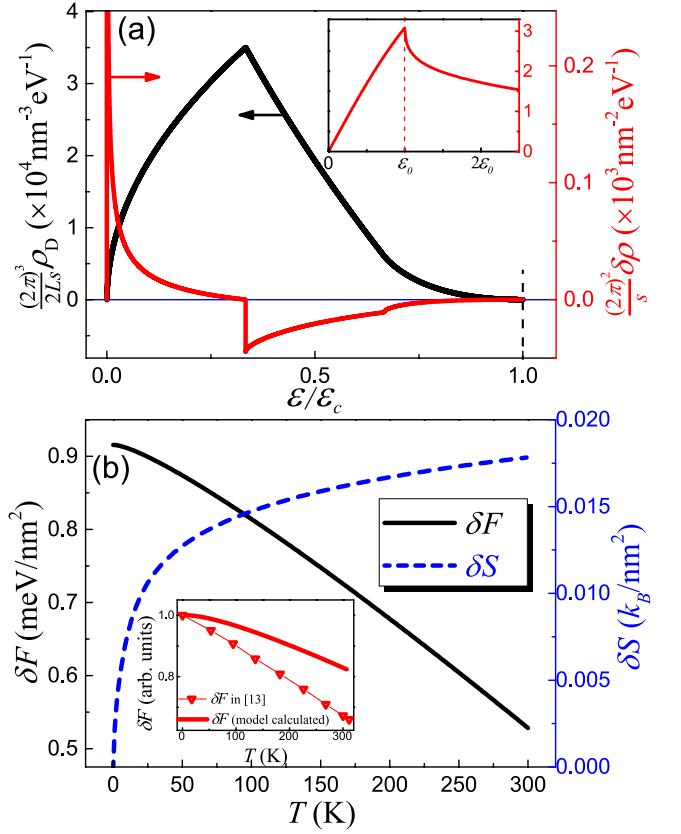}
\end{center}
\caption{(color online) (a) Magnon DOS of a domain $\frac{(2\pi)^3}{2Ls}
\rho_\mathrm{D}$ (black curve with left $y$-axis) and the DOS
difference $\frac{(2\pi)^2}{s}\delta\rho$ (red curve with right $y$-axis).
The black dashed line is $\varepsilon=\varepsilon_c$, and the red dashed
line is $\varepsilon=\varepsilon_0$. The inset is
$\frac{(2\pi)^2}{s}\delta\rho$ in $0\leq \varepsilon\leq 2.5\varepsilon_0$.
(b) The temperature dependence of the free energy density difference
$\delta{F}$ (black curve) and the entropy density difference $\delta{S}$
(blue curve) of a DW and a domain. Inset: Comparison of our analytical
$\delta F$ (solid line) with the numerical data in \cite{Nowak} (symbols) after calibrating the
0K energy. Only low temperature (below 311K that is much smaller than
than the Curie temperature) data is used where our model is justified.}
\label{fig2}
\end{figure}

Similarly, spin waves in a domain are plane waves whose dispersion
relation is the same as Eq. \eqref{dispersion} \cite{spinwave} with
the Brillouin zone $q_x,q_y,q_z\in[-\frac{\pi}{a},\frac{\pi}{a}]$.
The corresponding DOS is $\rho_\mathrm{D}({\varepsilon})=\int_{\mathrm{B.Z.}}
\frac{2Ls}{(2\pi)^3}\delta(\varepsilon-\hbar\omega_1)dq_xdq_ydq_z$.
The result of $\frac{(2\pi)^3}{2Ls}\rho_\mathrm{D}$ (black curve)
is shown in Fig. \ref{fig2}(a) (left $y$-axis)
with the YIG parameters \cite{Wangkl}: $A=2.48\times10
^{-16}$m$^2$, $M_s=0.84\times10^5$A/m, $\gamma=3.4\times10^4$Hz$
\cdot$m/A, $K_z=0.069$ and a shape anisotropy for a strip $K_x=1$.
The magnon DOS difference between a DW and a domain
$\frac{(2\pi)^2}{s}\delta\rho=\frac{(2\pi)^2}{s}(\rho_\mathrm{DW}-\rho_\mathrm{D})$
(red curve) for the same parameters is also plotted in Fig. \ref{fig2}(a)
(right $y$-axis). Below energy $\varepsilon_0$ there are only bound states
in a DW that do not exist in a domain so that $\delta\rho>0$
(shown in the inset for $0\le \varepsilon<2\varepsilon_0$).
The propagating states contribute to DOS for $\varepsilon\ge \varepsilon_0$.
The total area below the $\frac{(2\pi)^2}{s}\delta\rho$ curve is 0 because
the increase of DOS in low energy region comes from the decrease of DOS at
higher energy due to the phase shift of the propagating spin waves.
The total number of spin waves should be the same in a DW and in a domain.

The second law of thermodynamics says that a system should move toward
a state with a lower free energy or a larger entropy. Whether a DW should
move to the colder or the hotter regions of a wire under a temperature
gradient depends on the temperature dependence of the difference
of the DW free energy and the domain free energy, instead of DW
free energy only \cite{Nowak}. Let $U_0$ ($=4s\sqrt{AK_z}$
in our model) be the static DW energy (relative to that of a domain).
The energy of the DW with $\{n({\varepsilon})\}$ magnons of energy
${\varepsilon}$ is $E=U_0+\sum_{\varepsilon}n({\varepsilon}){\varepsilon}$.
The grand partition function $Z$ is $Z= \sum_{\{n({\varepsilon})\}}
e^{-\beta E}$, where the summation is over all possible magnon
configurations $\{n({\varepsilon})\}$, and $\beta=\frac{1}{k_B{T}}$
with $k_B$ the Boltzmann constant \cite{textbook}.
It is convenient to consider the free energy density per unit cross
section area which can be evaluated in a straightforward fashion.
The free energy density difference between a DW and a domain is thus
\setlength{\arraycolsep}{1pt}
\begin{eqnarray}
\delta{F}(T)&\equiv& F_\mathrm{DW}-F_\mathrm{D}=4\sqrt{AK_z}\notag\\
&+&\frac{k_B T}{s}\int_0^{{\varepsilon}_c}\ln(1-e^{-\beta {\varepsilon}})
\delta\rho({\varepsilon})d{\varepsilon}
\label{diff}
\end{eqnarray}
Notice that the magnon DOS difference between a DW and domain is
independent of $L$. Physically, this is because the free energy
difference between a DW and a domain should only relate to the
properties of a DW and a domain, characterized by $U_0$ and $\Delta$.

The black curve in Fig. 2(b) is the temperature dependence of $\delta F$
with the parameters mentioned above.
One can see that $\delta F$ always decreases with the increase of
the temperature, in an almost linear form as shown in the figure.
This behavior is directly related to the ``reshape" of magnon DOS:
Larger DOS at low energy means that the number of thermally excited
magnons is larger, this leads to a larger entropy. The blue line in
Fig. 2(b) is the temperature dependence of the density difference of the
DW entropy and the domain entropy, $\delta S=-\frac{\partial\delta F}
{\partial T}$. At 0K, only the lowest energy states (static DW or uniform
domain) are allowed for DWs or domains without any magnons. Thus the entropy
difference is zero and the free energy difference equals $U_\mathrm{DW}$.
As the temperature increases, the entropy of a DW is always larger
than that for a domain. Therefore, the free energy difference decreases
with $T$ monotonically as shown by the black line in Fig. 2(b).
Thus the total free energy can be lowered by moving the DW to
the hotter part of the wire. The basic thermodynamics principles
require a system to evolve in a way that lowers its free energy.
So as long as the spins interact with heat bathes this thermodynamic
force should always drive the DW to a well-defined direction--towards
the hotter part of the wire. The inset of Fig. 2(b) is the comparison
of our analytic result of $\delta F$ and the numerical result in
\cite{Nowak} with the same material parameters.
Results from two very different approaches compare well with each other
although our results involve many simplifications, including the exclusion
of the stray field due to the DW and simple cubic
crystal structure as well as temperature independence of model parameters.
In order to estimate the DW speed driven by a temperature gradient,
one can use $\delta{F}$ to find the equivalent magnetic field.
Then DW speed can be estimated by the Walker formula \cite{Walker,energy}.
Consider two points $A$ and $B$, which are $l$ apart from each other
along the wire, and assume $T_A$ and $T_B$ be the temperature at $A$
and $B$ with $T_B>T_A$. A DW moves from $A$ to $B$ if a DW is
initially centered at $A$. The free energy density of the
wire is then lowered by $\delta{F}(T_A)-\delta{F}(T_B)$.
Equating the decrease of this free energy with the
Zeeman energy by an equivalent magnetic field $H_{eq}$, one has
\begin{equation}
H_{eq}=\frac{\delta{F}(T_A)-\delta{F}(T_B)} {2\mu_0M_s l} = \frac{\delta{S}\nabla{T}}{2\mu_0M_s}.
\label{field}
\end{equation}
According to Eq. \eqref{diff} the equivalent field is independent
of sample size, $L$ and $s$, as expected.
Then according to the well-known Walker formula below the Walker
breakdown field $\alpha K_xM_s$, the DW speed $v$ under a field
$H_{eq}$ is $v=\gamma\frac{H_{eq}\Delta}{\alpha}$ \cite{Walker,energy}.
Together with Eq. \eqref{field} we can see the propagating speed $v$ is
proportional to the temperature gradient $\nabla{T}$. To compare with
the recent experiment \cite{Wangkl}, we use YIG parameters with $K_x=1$,
experimental temperature gradient of $\nabla{T}=2.25 \times10^4$K/m, and
$\alpha=0.0075$. Then the equivalent field is $H_{eq}\approx0.02$A/m.
The speed is then $v\sim7$mm/s which is about 1 order of magnitude
larger than the experimental value of $\sim200\mathrm{\mu}$m/s, a no
small value since modern laser technology can create a temperature
gradient as large as $10^9$K/m \cite{Nowak}. The discrepancy is not
surprising, considering complications involved in an experiment.
For a temperature gradient of $10^9$K/m, the effective equivalent field
$H_{eq}$ is about $2.7$mT which compares well with the estimated value
of $5$mT in Ref. \cite{Nowak} although the approach there is very
different from the current one.

There are fundamental differences in the STT interpretations
of magnon-driven DW motion and the thermodynamic viewpoint.
Magnonic STT can only predict DW motion correctly to a
system where the angular momentum dominates the DW dynamics.
However, the thermodynamic theory present here is general and
applicable to any wire with all possible microscopic interactions
as long as material parameters do not depend on the spin textures.
In case that material parameters depend on the spin textures, one
should expect very interesting and very rich physics. Of course, all
parameter changes should obey thermodynamic principles \cite{Bose}.
It is also clear that the thermodynamic theory is phenomenological in nature.
It provides no microscopic description of how spins interact with
other degrees of freedoms to generate a global DW propagation.
It should be pointed out that our theory considers only magnon
effects without electron contribution, important for a metallic wire.
It is known that electronic STT and magnonic STT have the opposite
sign under a temperature gradient. Thus, it is better to use magnetic
insulating wires if one wants to test the current theory so that
electron effects can be totally neglected. In a magnetic film,
magnetic domains form strips whose width decreases as the temperature
increase \cite{floating, domainsize,DWF}. It is interesting to note
that our theory can also provide a natural explanation to this
well-known fact: Because DW entropy is larger than that of a domain,
thus it is favorable to increase the number of DWs, or decrease strip
width in order to decrease the total free energy of the magnetic film.
The equilibrium value of the strip width is the compromise between
the entropy gain and energy cost in DW generation. Also, present
theory may also explain why skymions move to hotter region under a
thermal gradient if bound magnon states exist in skymions.
According to the present analysis,
the thermal gradient DW driven force is the entropy originated from
the magnon bound states. These states can also contribute to
the heat conduction in a DW \cite{Yanpeng}.

In conclusion, we compute magnonic contribution to the free energies
and entropies of a DW and a domain. It is analytically found, with
a clear physics picture, that a DW always has a larger entropy.
Thus, the driving force behind DW propagation under a temperature
gradient is the entropy. A DW propagates to the hotter region of a
wire in order to lower the wire free energy.
This result is robust and general. It does not depend on the microscopic
details of a wire or a DW as long as the material parameters
such as the exchange coefficient do not depend on the spin texture.
The DW propagating speed is proportional to temperature gradient and
can be as large as tens of m/s in reasonable parameters.
The free energy and/or entropy results can also explain the decrease
of domain size at a higher temperature.

This work was supported by Hong Kong RGC (Grants 604109 and
605413), and by NNSF of China (Grant 11374249).


\begin{thebibliography}{99}
\bibitem {Cowburn}D. A. Allwood, G. Xiong, C. C. Faulkner, D. Atkinson, D.
Petit, and R. P. Cowburn, Science \textbf{309}, 1688 (2005).
\bibitem {Parkin}S. S. P. Parkin, M. Hayashi, and L. Thomas, Science
\textbf{320}, 190 (2008).

\bibitem {Walker}N. L. Schryer and L. R. Walker,
J. Appl. Phys. \textbf{45}, 5406 (1974).
\bibitem {energy}X.R. Wang, P. Yan, J. Lu and C. He, Ann. Phys. (N.Y.)
\textbf{324}, 1815 (2009); X.R. Wang, P. Yan, and J. Lu,
Europhys. Lett. \textbf{86}, 67001 (2009).
\bibitem {WXS}X.S. Wang, P. Yan, Y.H. Shen, G. E.W. Bauer, and
X.R. Wang, Phys. Rev. Lett. \textbf{109}, 167209 (2012).

\bibitem {STT}L. Berger, Phys. Rev. B 54, 9353 (1996);J. Slonczewski, J.
Magn. Magn. Mater. 159, L1 (1996).
\bibitem {ono}A. Yamaguchi, T. Ono, S. Nasu, K. Miyake, K. Mibu and
T. Shinjo, Phys. Rev. Lett. \textbf{92}, 077205 (2004).
\bibitem {SZhang}S. Zhang and Z. Li, Phys. Rev. Lett.
\textbf{93}, 127204 (2004).

\bibitem {Bauer}M. Hatami, G. E.W. Bauer, Q. Zhang and P. J. Kelly,
Phys. Rev. Lett. \textbf{99}, 066603 (2007).

\bibitem {sse1}K. Uchida, S. Takahashi, K. Harii, J. Ieda, W. Koshibae,
K. Ando, S. Maekawa and E. Saitoh, Nature \text{455}, 778 (2008).
\bibitem {sse2}K. Uchida, J. Xiao, H. Adachi, J. Ohe, S. Takahashi, J. Ieda,
T. Ota, Y. Kajiwara, H. Umezawa, H. Kawai, G.E.W. Bauer,
S. Maekawa and E. Saitoh, Nat. Mater. \textbf{9}, 894 (2010).
\bibitem {sse3}H. Adachi, K. Uchida, E. Saitoh and S. Maekawa,
Rep. Prog. Phys. \textbf{76}, 036501 (2013).

\bibitem {Nowak}D. Hinzke and U. Nowak, Phys. Rev. Lett. \textbf{107}, 027205 (2011).
\bibitem {YT}A. A. Kovalev and Y. Tserkovnyak,
Europhys. Lett. \textbf{97}, 67002 (2012).
\bibitem {magnon}P. Yan, X.S. Wang, and X.R. Wang, Phys. Rev. Lett.
\textbf{107}, 177207 (2011).
\bibitem {Wangkl}W.J. Jiang, P. Upadhyaya, Y.B. Fan, J. Zhao,
M.S. Wang, L.-T. Chang, M.R. Lang, K.L. Wong, M. Lewis, Y.-T. Lin,
J.S. Tang, S. Cherepov, X.Z. Zhou, Y. Tserkovnyak, R. N. Schwartz,
and K.L. Wang, Phys. Rev. Lett. \textbf{110}, 177202 (2013)

\bibitem {Kim}D.S. Han, S.K. Kim, J.Y. Lee, S.J. Hermsoerfer, H. Schutheiss,
B. Leven, and B. Hillebrands, Appl. Phys. Lett. \textbf{94}, 112502 (2009).
\bibitem {Klaui}J.-S. Kim, M. St\"ark, M. Kl\"aui, J. Yoon, C.-Y. You,
L. Lopez-Diaz and E. Matrinez, Phys. Rev. B \textbf{85}, 174428 (2012).
\bibitem {Guo}X.-G. Wang, G.-H. Guo, Y.-Z. Nie, G.-F. Zhang, and Z.-X. Li,
Phys. Rev. B \textbf{86}, 054445 (2012).

\bibitem {Miron}I. M. Miron, T. Moore, H. Szambolics,
L.D. Buda-Prejbeanu, S. Auffret, B. Rodmacq, S. Pizzini,
J. Vogel, M. Bonfim, A. Schuhl and G. Gaudin,
Nat. Mater. \textbf{10}, 419 (2011).
\bibitem {Beach}S. Emori, U. Bauer, S.-M. Ahn, E. Martinez and
G. S. D. Beach, Nat. Mater. \textbf{12}, 611 (2013).
\bibitem {Parkin2}K.-S. Ryu, L. Thomas, S.-H. Yang and
S. S. P. Parkin, Nat. Nanotechnol. \textbf{8}, 527 (2013).

\bibitem {Hinzke}D. Hinzke, N. Kazantseva, U. Nowak, O. N. Mryasov,
P. Asselin, and R. W. Chantrell1, Phys. Rev. B \textbf{77}, 094407 (2008).

\bibitem {floating}O. Portmann, A. Vaterlaus, and D. Pescia,
Phys. Rev. Lett. \textbf{96}, 047212 (2006).
\bibitem {domainsize}N. Saratz, A. Lichtenberger, O. Portmann,
U. Ramsperger, A. Vindigni, and D. Pescia, Phys. Rev. Lett.
\textbf{104}, 077203 (2010).
\bibitem {DWF}B. Sangiorgio, T. C. T. Michauls, D. Pesia and A. Vindigni, arXiv:1308.3984.

\bibitem {LLG}T. L. Gilbert, IEEE. Trans. Magn. \textbf{40}, 3443 (2004).

\bibitem {YIG}A. A. Serga, A.V. Chumak and B. Hillebrands,
J. Phys. D: Appl. Phys. \textbf{43}, 264002 (2010).

\bibitem {spinwave}J. M. Winter, Phys. Rev. \textbf{124}, 452 (1961); A.A. Thiele, Phys. Rev. B \textbf{7}, 391 (1973).

\bibitem {textbook}L. D. Landau and E. M. Lifchitz,
\textit{Statistical Physics},(Pergamon Press 1969).

\bibitem {Bose}T. Bose and S. Trimper, Phys. Lett. A, \textbf{376}, 3386(2012).

\bibitem {Yanpeng}P. Yan and G. E.W. Bauer,
Phys. Rev. Lett. \textbf{109}, 087202 (2012).

\end{thebibliography}
\end{document}